\documentclass[aps,prl,amsmath,amssymb,twocolumn]{revtex4}
\usepackage{units}

\usepackage[]{graphicx}
\usepackage{times}
\usepackage{bm}
\usepackage{color}

\renewcommand{\bm }{\mathbf}

\newcommand{\comment}[1]{}

\begin{document}

\title{Mapping of the dark exciton landscape in transition metal dichalcogenides}
\author{Gunnar Bergh\"auser$^{1}$}
\email{gunbergh@chalmers.se}
\author{ Philipp Steinleitner$^{2}$, Philipp Merkl$^{2}$, Rupert Huber$^{2}$, Andreas Knorr$^3$, and Ermin Malic$^1$}

\affiliation{$1$ Department of Physics, Chalmers University of Technology, Gothenburg, Sweden, $2$
Department of Physics, University of Regensburg, 93040 Regensburg, Germany,
$3$ Institut f\"ur Theoretische Physik, Technische Universit\"at Berlin, 10623 Berlin, Germany}
\begin{abstract}
Transition metal dichalcogenides (TMDs) exhibit a remarkable exciton physics including optically accessible (bright) as well as spin- and momentum-forbidden (dark) excitonic states. 
So far the dark exciton landscape has not been revealed leaving in particular the spectral position of momentum-forbidden dark states completely unclear. 
This has a significant impact on the technological application potential of TMDs, since the nature of the energetically lowest state determines, if the material is a direct-gap semiconductor.
Here, we show how dark states can be experimentally revealed by probing the intra-excitonic 1s-2p transition. Distinguishing the optical response shortly after the excitation ($< \unit[100]{fs}$) and after the exciton thermalization ($> \unit[1]{ps}$) allows us to demonstrate the relative position of bright and dark excitons. 
We find both in theory and experiment a clear blue-shift in the optical response demonstrating for the first time the transition of bright exciton populations into lower lying momentum- and spin-forbidden dark excitonic states in monolayer WSe$_2$.

\end{abstract}
\maketitle

\begin{figure}[t!]
\begin{center}
\includegraphics[width=1.\linewidth]{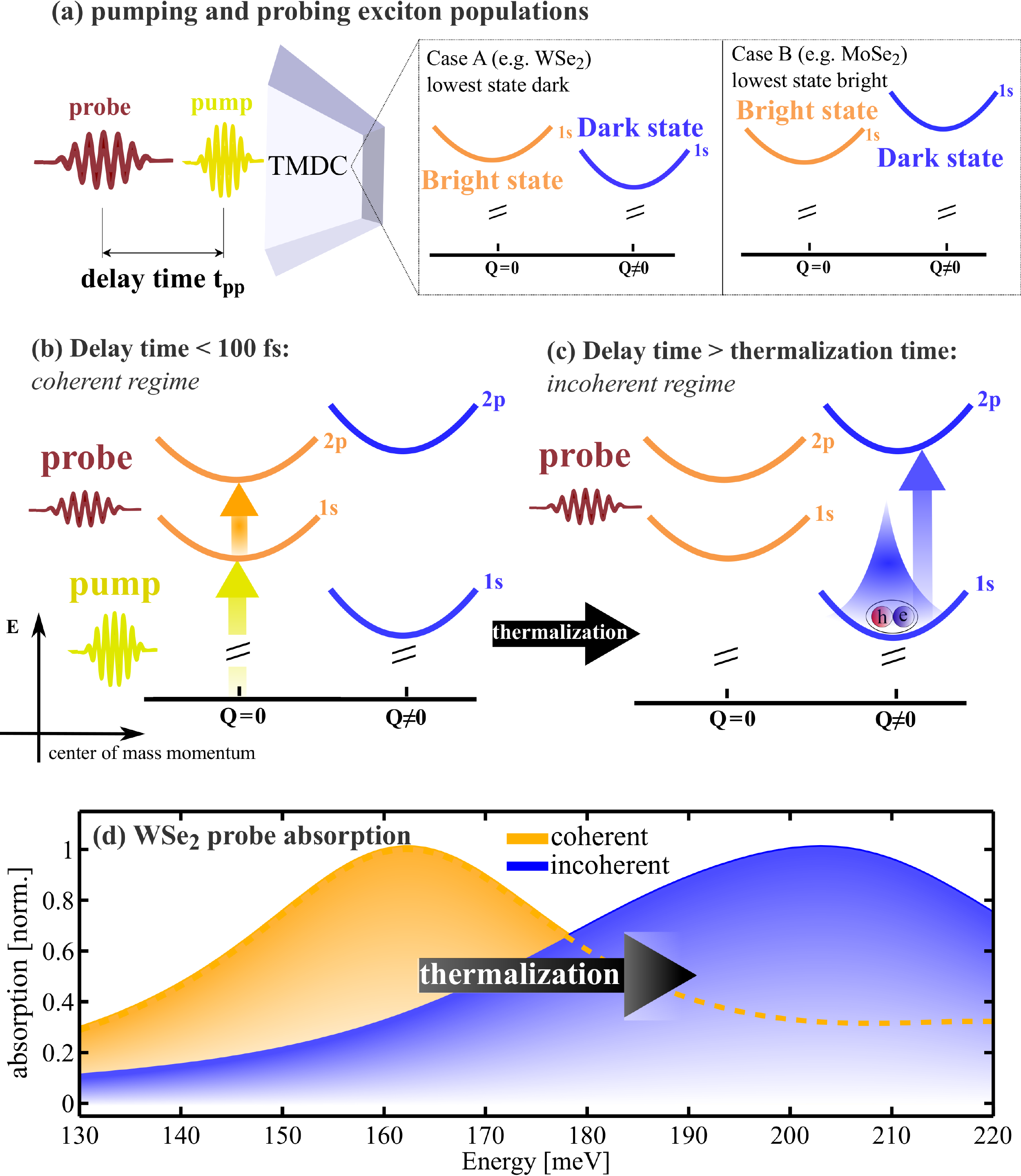}
 \end{center}
\caption{\textbf{Schematic illustration and theoretical prediction of infrared response.} (a) Schematic representation of the modeled experimental situation, where first a pump pulse induces a coherent exciton density in the 1s state, which is then probed with a weak infrared pulse with an energy matching the intra-excitonic 1s-2p transition. Here, we can distinguish between two scenarios, where the energetically lowest state is dark (e.g. K-$\Lambda$ exciton) or bright (K-K exciton).  (b) For very short delay times between the pump and the probe pulse (coherent limit), the coherent exciton density has not decayed yet and dominates the linear optical response.  (c) For longer delay times, incoherent excitons have already been formed and have thermalized through exciton-phonon coupling. In this incoherent limit,  the  optical response is dominated by the lowest lying excitonic states, where the population of incoherent excitons is the largest. (d) Absorption spectrum for WSe$_2$ in the coherent (yellow area) and incoherent (blue area) regime. The first reflects the population of bright K-K excitons, while the second results from the energetically lowest dark K-$\Lambda$ excitons.
}
\label{fig:1}
\end{figure}

Transition metal dichalcogenides (TMDs)  are characterized by tightly Coulomb-bound electron-hole pairs that efficiently  couple to light and that can be spin- 
and valley-selectively excited \cite{novoselov05, HeinzPRL2010, Splendiani2010,Chernikov2014}. Recent experimental and theoretical studies show that besides these bright excitonic states, dark optically inaccessible excitons \cite{Hongkun2017,Potemski2017a,2017_Heinz_nnano,MacDonald2015PRB,Berkelbach2015PRB,Malte2016a,Danovich2017a,2015_Steinleitner_NatMat}
play a significant role for the optical response \cite{2015_Heinz_PRL,Malte2016a,2017_Feierabend_NatComm} and the non-equilibrium dynamics in TMDs \cite{2017_Malte_Arxiv,Steinhoff2017a,Danovich2017a,Steinhoff2014}. 
In particular, the existence of energetically lower dark exciton states strongly limits the photoluminescence quantum efficiency of these materials \cite{2015_Heinz_PRL,2017_Malte_Arxiv,Darexcitonpaper}. 
We distinguish spin-forbidden dark states, where the Coulomb-bound electron and hole exhibit opposite spin, from momentum-forbidden dark states exhibiting a non-zero center-of-mass or angular momentum. 
Both cannot be accessed by light, since photons cannot provide the required spin or momentum to induce an interband transition into these states. 
Exciton populations occupying dark states can be optically addressed using mid-infrared spectroscopy \cite{2015_Steinleitner_NatMat,2017_Steinleitner_NanoLetter} 
, however so far, the focus of these studies has been on revealing the radiative recombination time and the exciton formation dynamics.  The exciton landscape and the microscopic origin of different types of  dark excitonic states has not been addressed yet.
In particular, the spectral position of momentum-forbidden dark excitons has not been revealed, leaving the nature of the energetically lowest states unclear. 
This is of crucial importance for the technological application potential of TMDs, since it determines whether the material is a direct or an indirect-gap semiconductor.\\
Here, we show how the entire exciton landscape including spin- and momentum-forbidden dark excitons can be identified. We combine a sophisticated  microscopic theory \cite{2008_Steiner_PRL,2008_Steiner_PRB,2010_Steiner_PRL} with femtosecond infrared spectroscopy\cite{2016_Boland_NNano,2017_Jigang_Nature,2015_Steinleitner_NatMat,2017_Steinleitner_NanoLetter} probing the intra-excitonic transition between 1s and 2p states of different exciton types, cf. Figs. \ref{fig:1}(a)-(b).

While the 1s-2p transition is in the THz range in conventional semiconductors \cite{2008_Steiner_PRL,2008_Steiner_PRB,2010_Steiner_PRL,Kida2005,2007_Masayoshi_NatPhot,2017_Jigang_Nature}, the excitonic binding energies in TMDs are one order of magnitude larger and the probing of exciton populations is therefore in the mid-infrared regime \cite{2017_Steinleitner_NanoLetter,2015_Steinleitner_NatMat,2016_Hyunyong_NatCom}. 
Each excitonic state is characterized by its binding energy, which also determines the 1s-2p transition energy. As a result, we can identify different excitonic states based on their linear response to an infrared probe pulse. 
Furthermore, distinguishing the  response shortly after the optical excitation (coherent regime, $< \unit[100]{fs}$) and after the
exciton thermalization (incoherent regime, $> \unit[1]{ps}$), we can track the exciton dynamics shifting the population from optically excited coherent excitons to incoherent excitonic states that are formed via exciton-phonon scattering 
\cite{2017_Malte_Arxiv}. After thermalization, the largest exciton population will be accumulated in the energetically lowest state according to the equilibrium Bose distribution. While the coherent response is always determined by the 1s-2p transition of the bright exciton with zero center-of-mass momentum $Q$, the incoherent response will be dominated by the energetically lowest exciton (Figs. \ref{fig:1}(b)-(c)). The latter is characterized by a  larger excitonic binding energy and thus a higher 1s-2p separation. 
 As a result, we expect a  blue-shift in the incoherent regime approximately $\unit[1-2]{ps}$ after the optical excitation for materials with dark excitons as the energetically lowest states. In this work, we demonstrate both in theory and experiment a clear blue-shift in the optical response (Fig. \ref{fig:1}(d)) demonstrating for the first time the existence of energetically lower lying momentum-forbidden dark excitonic states in monolayer tungsten diselenide (WSe$_2$).

To map the excitonic landscape in TMDs, we perform femtosecond infrared spectroscopy experiments and sophisticated calculations on microscopic footing based on the density matrix formalism \cite{PQE,Kochbuch, malic13}  and TMD Bloch equations \cite{Bergh2014b,Malte2016a, 2017_Feierabend_NatComm}.
 To obtain access to the optical response of TMD materials to an infrared pulse, we first solve the Heisenberg equation of motion for the microscopic polarization $p_{ij}(t)=\langle a_i^\dagger a_j \rangle (t)$ with the fermionic creation and annihilation operators $a_i^\dagger$ and $a_j$ and the compund indices $i,j$ containing all electronic quantum numbers, such as the band index $\lambda$ and the  momentum $\bm q$. The microscopic polarization presents a measure for optical inter- and intra-band transitions. 
To account for the strong exciton physics in TMDs, this microscopic quantity is projected to an excitonic basis with $p_{\mathbf Q}^{n}=\sum_{\mathbf q}p_{\mathbf q,\mathbf Q}^{cv}\phi_{\mathbf q}^{n}$   \cite{PQE,Kochbuch,Bergh2014b}.
The appearing  eigenfunctions $\phi_{\mathbf q}^{n}$  are obtained by solving  the Wannier equation, which is an eigenvalue equation for the exciton with the index $n$. Here, we take into account optically accessible bright  excitons (K-K excitons) that are located  within the light cone characterized by the center-of-mass momentum $\bf Q\approx 0$. Furthermore, we include momentum-forbidden intervalley dark excitonic states, where the Coulomb-bound hole and electron are located in different valleys. In particular, we include K-$\Lambda^{(\prime)}$ and K-K$^{(\prime)}$ ($\Gamma-\Lambda^{(\prime)}$ and $\Gamma$-K$^{(\prime)}$) states with the hole at the K ($\Gamma$) point and the electron either in the $\Lambda^{(\prime)}$ or the K$^{(\prime)}$ valley. Here, we also take into account corresponding spin-forbidden excitonic states, where the Coulomb-bound electron and hole show opposite spin.

The goal is to calculate the optical response of different monolayer TMDs to a weak infrared probe pulse after coherently pumping the 1s exciton in two limiting cases: (i) In the coherent limit shortly  after the pump pulse (delay time $\text{t}_{\text{pp}} < \unit[100]{fs}$), when the system is dominated by the optically generated excitonic coherence $p_{\mathbf Q}^{n}$, cf. the orange line in Fig. \ref{fig:1} (d).  (ii) In the incoherent limit after exciton thermalization (delay time $\text{t}_{\text{pp}} > \unit[1]{ps}$), when all interband coherences have decayed and the optical response is solely determined by the incoherent exciton density $N_{\mathbf Q}^n$ reflecting an equilibrium Bose distribution (cf. the blue line in  Fig. \ref{fig:1} (d)). While the pump pulse excites coherent excitons, e.g. creates an excitonic polarization, the infrared probe pulse induces intra-excitonic coherences inducing transitions from  1s to 2p excitonic states (Fig. \ref{fig:1} 
(c)). To determine the infrared absorption of TMDs, we  calculate the macroscopic intraband current, which reads in the excitonic picture 
\begin{align}
\label{eq:MacrointrabandCurrenttime}
J (t)=
 \sum_{nm}\sum_{\mathbf Q}j_{nm}\bigg(
  p^{n*}_{\mathbf Q}(t)p^m_{\mathbf Q}(t)
   + N_{\mathbf Q}^{nm}(t)\bigg).
\end{align}
This equation contains both the infrared response in the coherent (first term) and incoherent limit (second term). Here,  we have introduced the excitonic intraband current $j_{nm}=\sum_{\mathbf q,\lambda}j_{\lambda}(\mathbf{q}) \phi^{n*}_{\mathbf q}\phi^{m}_{\mathbf q}$ corresponding to the current $j_\lambda(\bf q)=-\frac{e_0}{m_\lambda}\bm k$ weighted with excitonic wave functions. Here, $m_\lambda$ corresponds to the effective mass of the considered electronic band $\lambda$. Furthermore,  $N_{\mathbf Q}^{nm}$ describes the incoherent exciton density for $n=m$, while in the considered thermalized equilibrium situation the off-diagonal terms $n\neq m$ vanish.
The dynamics of the excitonic polarization $p^n_{\mathbf Q}(t)$ as a response to a weak probe pulse is given by the TMD Bloch equation, cf. Eq. (\ref{eq:dynexc}) in the appendix.

The exciton distribution $N^n_{\mathbf Q}$(t) can be obtained by taking into account the phonon-assisted formation of incoherent excitons as well as their thermalization towards an equilibrium distribution. While in a previous work \cite{2017_Malte_Arxiv} we have performed the full time- and momentum-dependent calculations, here we exploit the obtained thermalized Bose distribution of incoherent excitons. After solving  the TMD Bloch equation for the excitonic polarization $p_{\mathbf Q}^{n}(t)$ in the coherent limit, we have access to the intra-band current from  Eq.(\ref{eq:MacrointrabandCurrenttime}). 
Then, we can determine the optical susceptibility 
  $\chi(\omega)=\frac{J(\omega)}{ \varepsilon_0 \omega^2 A_{0}(\omega)}$, where $\varepsilon_0$
  is the vacuum permittivity.
  Assuming an ultrashort and weak intraband probe pulse and furthermore neglecting terms higher than the 4th order in the field, the linear mid-infrared susceptibility reads in the coherent ($\chi^c$) and incoherent limit ($\chi^{ic}$)  
\begin{align}
\label{eq:ChiCo}
\chi^c(\omega)&=
\frac{1}{\varepsilon_0\omega^2}\sum_{nm}\left(
\frac{
|j_{1s,m}|\,|p^{n}_{0}|^2
}{\Delta \varepsilon_{m,n}-i\gamma_{0,c}^{mn}-\hbar \omega}+ c.c.\right),\\[8pt]\label{eq:ChiInco}
  \chi^{ic}(\omega)&=
  \frac{1}{\varepsilon_0\omega^2}
  \sum_{\mathbf Q nm} j_{nm}
  \frac{\sum_{l}\left(
  j_{ml} N_{\mathbf Q}^{nl}
 -j_{ln} N_{\mathbf Q}^{lm} \right)}{\varepsilon^{m}_{\mathbf Q}-\varepsilon^{n}_{\mathbf Q}-i\gamma_{\mathbf Q,ic}^{mn}-\hbar\omega}.
\end{align}
Here, we have introduced the quasi-static pump-induced excitonic polarization $p^{n}_{0}$ at the time of the probe pulse
and a constant dephasing rate for the intra excitonic transition including $\gamma^{mn}_{0,c}=\gamma^{n}_{0,c}+\gamma^{m}_{0,c}$
in the coherent and  $\gamma^{mn}_{\mathbf Q,ic}$ in the incoherent regime. 
 Since a microscopic description of the dephasing rates stemming from higher-order correlations is beyond the scope of this work, we assume a typical constant dephasing of 40 meV \cite{Malte2016a}.

In both  regimes, we find that the absorption shows a Lorentzian shape. The spectral position of the Lorentzians is determined by the energy difference of the initial (n) and the final (m) excitonic state. The absorption intensity is given by the optically excited coherent or indirectly formed incoherent exciton populations in the involved states. While the population of the initial state enhances the absorption of the mid-infrared probe pulse, the population of the final state leads to negative contributions that could eventually result in gain \cite{2004_PRL_KochThZGain}. 

\begin{figure}[t!]
\begin{center}
\includegraphics[width=1\linewidth]{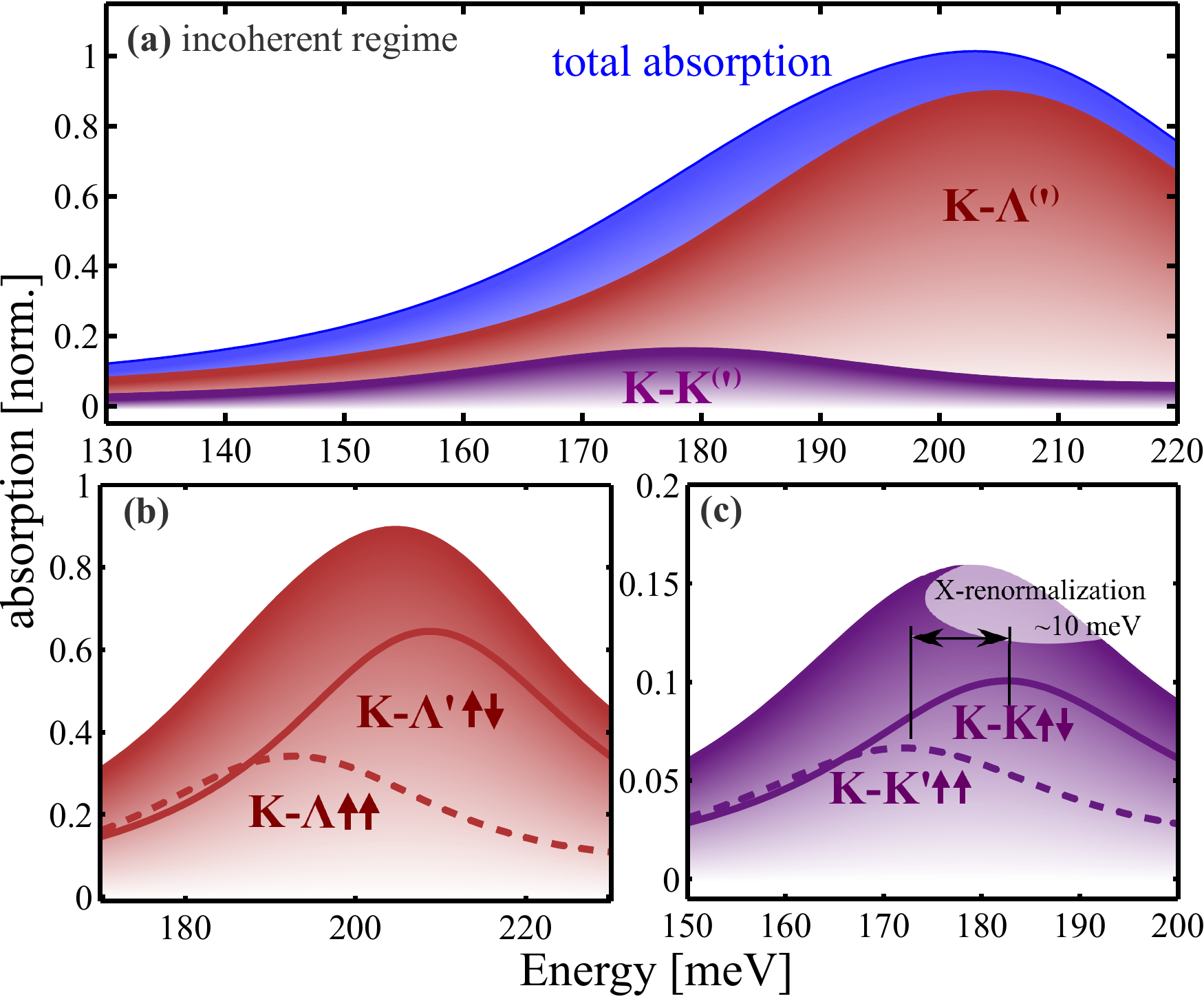}
 \end{center}
\caption{\textbf{Contribution of different excitonic states.}
(a) Linear optical response after exciton thermalization (incoherent regime) in WSe$_2$. 
The largest absorption intensity stems by far from the energetically  lowest K-$\Lambda^{(\prime)}$ excitons (red line) followed by  K-K$^{(\prime)}$ excitonic states (purple line). 
Further resolution of the contribution of these states includes (b) spin-like K-$\Lambda$ ($\uparrow\uparrow$) and spin-unlike K-$\Lambda^{\prime}$ ($\uparrow\downarrow$) excitons and (c) spin-like K-K' and spin-unlike K-K excitons, where the Coulomb-bound electron and hole exhibit either the same or the opposite spin. 
}
\label{fig:2}
\end{figure}

Evaluating Eqs. (\ref{eq:ChiCo}) and (\ref{eq:ChiInco}), we have access to the optical response of TMD materials to an infrared probe pulse both in the coherent regime ($\text{t}_{\text{pp}}<\unit[100]{fs}$) directly after the pump pulse and in the incoherent regime after the excitonic thermalization ($\text{t}_{\text{pp}}>\unit[1]{ps}$). The corresponding probe absorption is shown in Fig. \ref{fig:1} (d) for the exemplary material tungsten diselenide (WSe$_2$). 
In the coherent limit (yellow area), the initially pumped coherent 1s excitons 
dominate the optical response to the infrared pulse. Here, the pronounced peak is located at $\unit[160]{meV}$ corresponding to the energy difference between the 1s and the 2p excitonic state of the bright K-K exciton. 
Note  that the  2p exciton consists of the states 2p$_{+}$ and 2p$_-$, which  are not energetically degenerated due to a non-vanishing geometric phase in TMD monolayers \cite{2016_bergh_2Dmat}. As a result, we find that the infrared absorption is given by a a superposition of two Lorentzian peaks. 

For larger delay times between the pump and the probe pulse,  incoherent exciton populations are formed \cite{Malte2016a,2017_Malte_Arxiv} and have already thermalized into a Bose distribution. Here, predominantly lower energetic states are occupied, cf. Fig. \ref{fig:1} (c). The excitonic landscape of WSe$_2$  exhibits optically inaccessible dark states well below the initially pumped bright K-K exciton \cite{2017_Malte_Arxiv}. 
These momentum-forbidden intervalley excitons (K-$\Lambda$) are located at different high-symmetry points of the Brillouin zone. The involved conduction band at the $\Lambda$ valley shows a significantly larger effective mass compared to  the K valley \cite{Kormanyos2014} resulting in a higher excitonic binding energy and therefore a larger 1s-2p separation. In the incoherent regime, we find that the infrared resonance is blue-shifted by approximately $\unit[40]{meV}$, cf. the blue area in Fig. \ref{fig:1}(d). The new peak  at $\unit[200]{meV}$ corresponds to the  1s-2p transition energy of the $K-\Lambda$ exciton.

Investigating in more detail the contribution of different excitonic states to the optical response in the incoherent limit, we find that  the dark K-$\Lambda^{(\prime)}$ excitons clearly dominate followed by the contribution of K-K$^{(\prime)}$ excitons giving rise to a shoulder at energies around $\unit[180]{meV}$, cf. Fig. \ref{fig:2}(a). 
Our approach allows to further resolve the contribution of spin-unlike (spin-forbidden, $\uparrow\downarrow$) and spin-like (spin-allowed, $\uparrow\downarrow$) K-$\Lambda^{(\prime)}$ and K-K$^{(\prime)}$ states. Figures \ref{fig:2}(b)-(c) show that by far the most pronounced optical response stems from the spin-unlike K-$\Lambda'$ exciton, which is the energetically lowest state. In contrast, the spin-like  K-$\Lambda$ exciton is far above the bright state due to the large spin-orbit splitting and therefore only marginally contributes (not shown). We also find a significant contribution from the spin-unlike (around $\unit[205]{meV}$) and the spin-like (around $\unit[190]{ meV}$) K-$\Lambda$ exciton followed by the spin-unlike (around $\unit[180]{ meV}$) K-K and the spin-like (around $\unit[170]{ meV}$) K-K' excitons (Fig. \ref{fig:2} (c)).
 The energetic separation of these states can be explained by the different energy renormalization stemming from the Coulomb exchange coupling (X renormalization). The latter gives rise to a shift of spin-like 1s excitonic states, while it vanishes for spin-unlike states and for 2p excitons due to symmetry reasons \cite{Barry_PRL_Louie2015,Echeverry2016PRB,MacDonald2015PRB}.

\begin{figure}[t!]
\begin{center}
\includegraphics[width=0.85\linewidth]{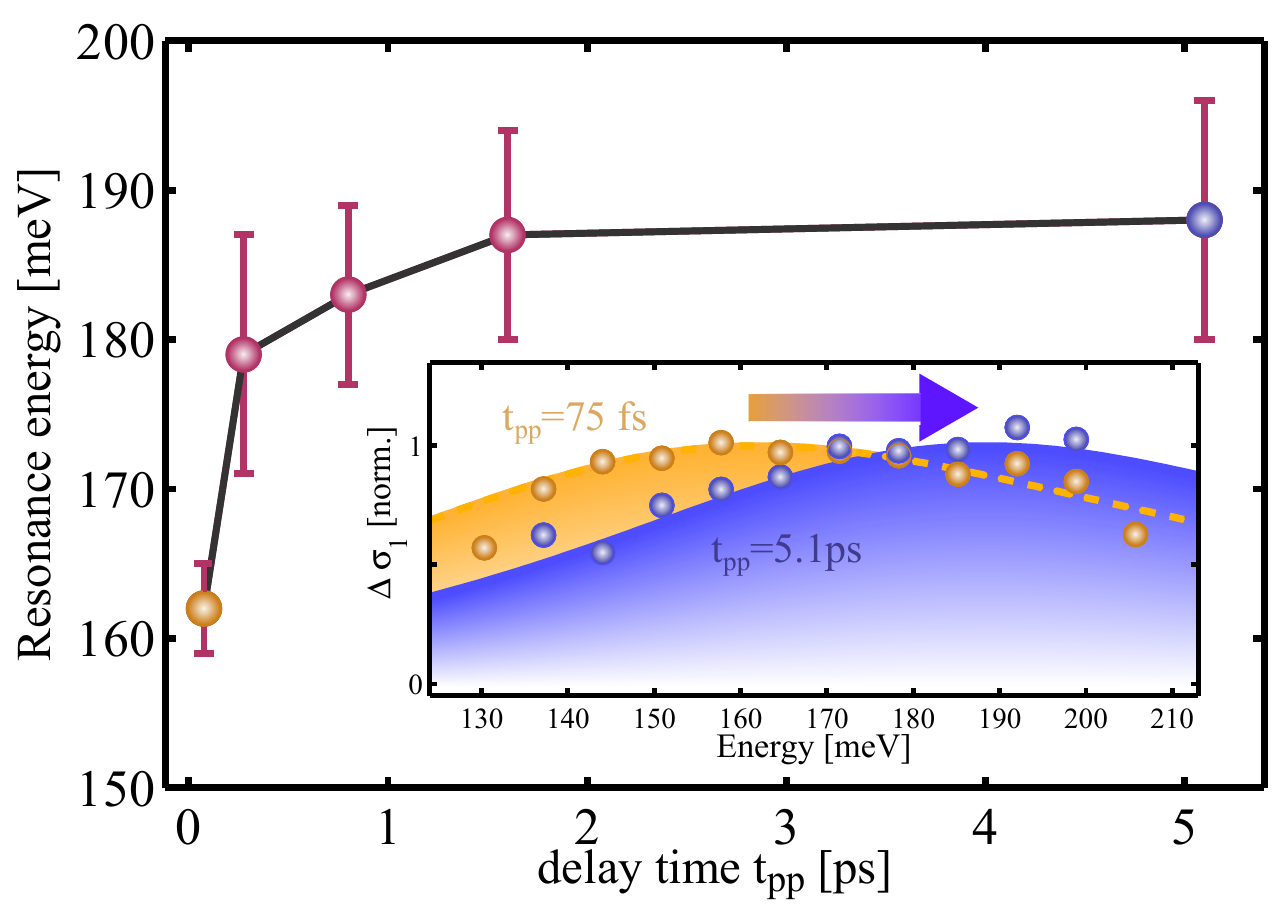}
 \end{center}
\caption{\textbf{Experimentally measured blue-shift}. Resonance energy of the intra-excitonic 1s-2p transition in a monolayer WSe$_2$ as a function of the delay time $\text{t}_{\text{pp}}$ after resonant interband optical injection of coherent 1s A excitons.  The error bars represent the 95\% confidence interval of the resonance energy. The inset shows the real part of the pump-induced mid-infrared conductivity $\Delta \sigma_1$ (corresponding to the optical absorption) as a function of the photon energy for two characteristic pump delay times $\text{t}_{\text{pp}}$. Here, colored spheres correspond to experimental data, while shaded areas show the result of the Lorentz-oscillator model  \cite{2015_Steinleitner_NatMat} fitting $\Delta \sigma_1$ and allowing for the extraction of the resonance energy depicted in the main figure.
}
\label{fig:3}
\end{figure}

In order to test the theoretical predictions, we perform a femtosecond near-infrared pump / mid-infrared probe experiment in a WSe$_2$ monolayer: A 90-fs laser pulse centered at a wavelength of $\unit[742]{nm}$ selectively injects bright K-K excitons with a vanishing center-of-mass momentum ($\bm Q \approx 0$). At a variable delay time $\text{t}_{\text{pp}}$, we subsequently probe the 1s-2p intra-excitonic transition with phase-locked few-cycle mid-infrared pulses covering the range of photon energies between $\unit[125]{meV}$ and $\unit[210]{meV}$. Complete amplitude- and phase-sensitive electro-optic detection of the transmitted probe waveform allows us to extract the full complex-valued response function of the exciton ensemble \cite{2015_Steinleitner_NatMat}. The inset of Fig. \ref{fig:3} depicts the real part of the pump-induced mid-infrared conductivity $\Delta\sigma_1$ (corresponding to optical absorption) for two characteristic delay times $\text{t}_{\text{pp}}$. The pump fluence $\Phi = \unit[16]{\mu J/cm^2}$ is chosen such that a moderate exciton density of approximately $\unit[10^{12}]{cm^{-2}}$ is maintained. For both delay times, a clear peak in $\Delta\sigma_1$ is observable corresponding to the absorptive intra-excitonic 1s-2p transition \cite{2015_Steinleitner_NatMat}. The peak energy of this transition exhibits a distinct blue-shift as the delay time increases from $\text{t}_{\text{pp}} = \unit[75]{fs}$ to $\text{t}_{\text{pp}} = \unit[5.1]{ps}$. A more systematic study of the ultrafast evolution of the resonance energy extracted by fitting the experimental data with a Lorentz-oscillator is shown in the main Fig. \ref{fig:3}. The resonance energy shows a strong blue shift from $\unit[162]{meV}$ at $\text{t}_{\text{pp}} = \unit[75]{fs}$ to $\unit[179]{meV}$ at $\text{t}_{\text{pp}} = \unit[275]{fs}$. Within the next few picoseconds the resonance energy still slightly increases, leading to a saturation at a delay time $\text{t}_{\text{pp}} = \unit[5.1]{ps}$ and a total blue shift of $\unit[26]{meV}$. This observation is in excellent agreement with the theoretical prediction of a blue-shift during the process of exciton thermalization resulting in a strong occupation of  energetically lowest dark excitonic states. Once the thermalization is reached after approximately $\unit[2]{ps}$ \cite{2017_Malte_Arxiv}, the blue-shift is expected to saturate, as observed in the experiment. 

\begin{figure}[t!]
\begin{center}
\includegraphics[width=1\linewidth]{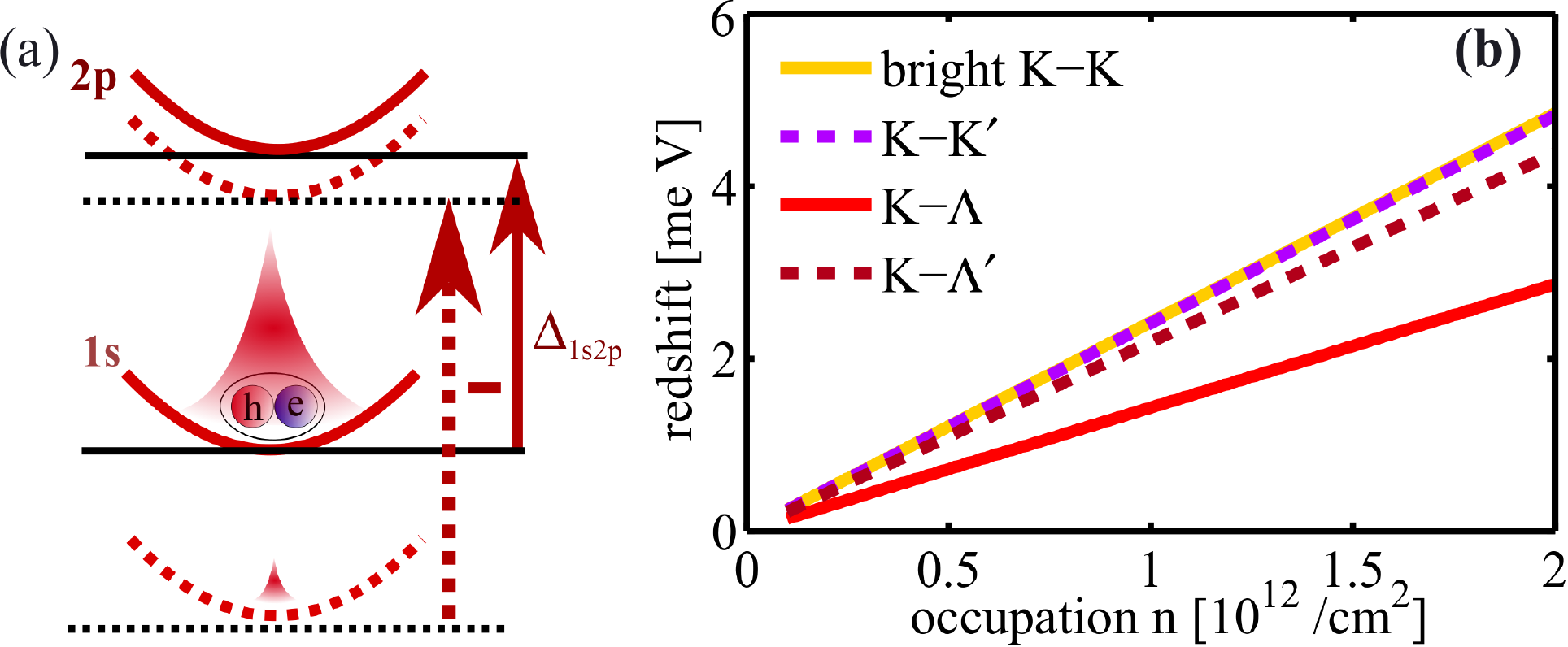}
 \end{center}
\caption{ \textbf{Population-induced spectral shift.}
(a) Schematic illustration of the red-shift of the 1s-2p transition $\Delta_{1s, 2p}= E_{2p}-E_{1s}$ for increasing  exciton population (dashed to solid lines), which can be ascribed to the larger reduction of the 1s state energy. (b) The predicted  red-shift  for the bright K-K exciton (yellow) as well as the dark K-K' (purple dashed), K-$\Lambda$  (red) and K-$\Lambda^\prime$ excitons (dark red dashed) as a function of the exciton population.
}
\label{fig:4}
\end{figure}

To make sure that the observed shift is not a result of a population-induced energy renormalization, we explicitly calculate the shift of the involved bright K-K as well as dark K-K', K-$\Lambda$, and K-$\Lambda^\prime$ excitons as a function of the optically excited carrier occupation.
The latter leads to an enhanced screening of the Coulomb potential and to Pauli blocking effects that reduce excitonic binding energies and consequently influence the 1s-2p transition, cf. Fig. \ref{fig:4}(a). These effects are accounted for dynamically in the $\Gamma$ term appearing in  Eq.(\ref{eq:dynexc}) in the appendix. 
In Fig. \ref{fig:4}(b), we show the  dependence of the relative shift of 1s and 2p states with the carrier population.
 Increasing the population results in a reduced excitonic binding energy, which leads to red-shift of the 1s-2p transition. In the modelled experimental situation, a pump pulse generates a coherent excitonic population that decays radiative and non-radiative. The decay reduces the population and therefore should lead to a blue-shift of the infrared resonance. One may therefore argue that the experimentally observed blue-shift at larger delay times  could be ascribed to this population-induced renormalization effect. 
 However, in a moderate population regime of approximately $\unit[10^{12}]{cm^{-2}}$ (well below the Mott transition \cite{Steinhoff2017a}) as discussed in the experiment, we find only minor shifts of 1 or $\unit[2]{meV}$ for a population change of $\unit[10^{12}]{cm^{-2}}$ confirming the trend measured in gating experiments \cite{2015_Chernikov_PRL} and in the pump intensity study of Ref. \onlinecite{2015_Steinleitner_NatMat}. As a result, the population-induced energy renormalization cannot explain the observed blue shift in the range of $\unit[30]{meV}$.

\begin{figure}[t!]
\begin{center}
\includegraphics[width=1\linewidth]{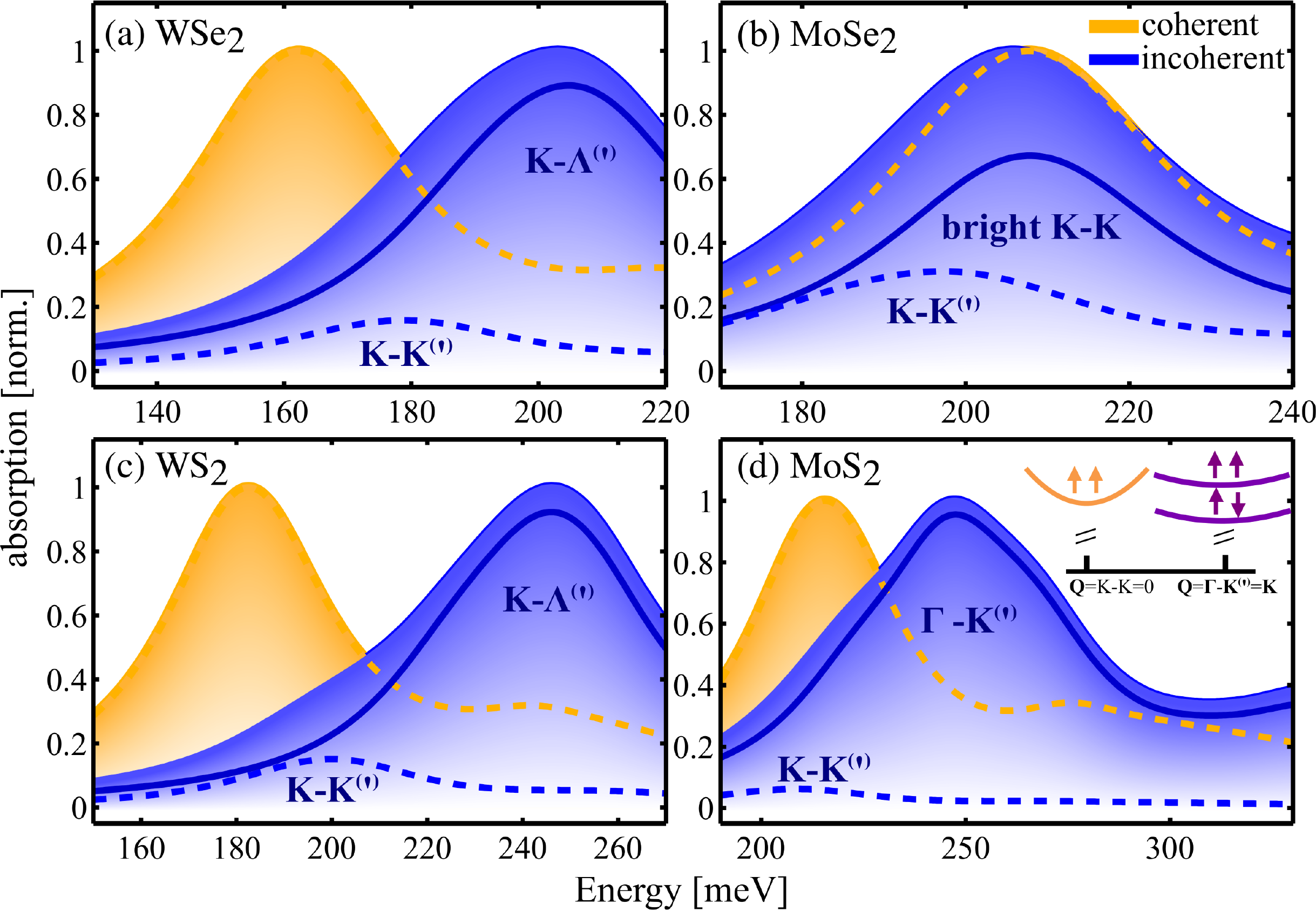}
 \end{center}
\caption{\textbf{Comparison of different TMDs.}
Optical response to an infrared red probe pulse for monolayer (a) WSe$2$, (b) MoSe$_2$, (c) WS$_2$, and (d) MoS$2$ distinguishing the coherent (orange) and incoherent (blue) contribution. One can directly read off, which TMDs are indirect-gap semiconductors, i.e. exhibit energetically lower lying dark excitonic states. The latter is demonstrated in a clear blue-shift of the incoherent contribution reflecting the higher excitonic binding energy and a larger 1s to 2p separation of the dark states. (a), (c) In W-based TMDs, thermal distribution of excitonic occupation favors the low-lying K-$\Lambda^{(\prime)}$ (solid blue lines) and K-K$^{(\prime)}$ states (dashed lines). (b) In contrast  the major contribution in MoSe$_2$ stems from the bright K-K exciton revealing the direct-gap nature of this material. (d) In MoS$_2$, the lowest states are the dark $\Gamma$-K$^{(\prime)}$ excitons, which dominate the optical response in the incoherent regime leading to a broad and blue-shifted resonance after exciton thermalization. 
}
\label{fig:5}
\end{figure}

The excitonic landscape of each TMD material is different and has been controversially discussed in literature. Although there is a huge number of investigations e.g. on  MoS$_2$, it is still unclear, whether this material is a direct or indirect-gap semiconductor. While a temperature-dependent PL study suggests it to be a direct semiconductor \cite{2015_Heinz_PRL} recent experimental data in an in-plane magnetic field imply that there should be a  lower lying spin-forbidden excitonic state \cite{Potemski2017a}.  Even theoretical calculations yield different ordering of dark and bright states in different TMD materials \cite{Barry_PRL_Louie2015,Echeverry2016PRB,MacDonald2015PRB,Steinhoff2017a}.
Our approach allows to directly address this question, since the optical infrared response significantly differs depending on the relative position of bright and dark excitonic states. While for direct semiconductors, only moderate population-induced blue shifts of the 1s-2p transition energy  are expected,  for indirect semiconductors these shifts will be in the range of a few tens of meV reflecting the relative position of dark excitons.
In Fig. \ref{fig:5}, we show a direct comparison of the infrared absorption of the four most studied TMD materials including (a) WSe$_2$,
(b) MoSe$_2$, (c) WS$_2$ and (d) MoS$_2$. 
Based on DFT-input parameters on the electronic effective masses and band gap energies  \cite{Kormanyos2014}, we find that  MoSe$_2$ is the only material that does not exhibit a blue-shift in the infrared response, cf. Fig\ref{fig:5} (b). Surprisingly, we observe a small red-shift due to a minor population of spin-unlike K-K and spin-like K-K' excitons (dashed line Fig\ref{fig:5} (b)). In contrast, the infrared absorption of all other studied TMDs exhibits pronounced blue-shifts in the range of $\unit[30-40]{meV}$ suggesting the existence of an energetically lowest dark state.
The dark excitonic landscape of WS$_2$ results in a similar infrared absorption as  already discussed in the case of  WSe$_2$, however exhibiting a larger blue-shift and a clearer separation of the contribution of K-K and K-$\Lambda$ excitons due to the larger effective masses of the $\Lambda$ valley  in WS$_2$ \cite{Kormanyos2014}.
For MoS$_2$, we reveal that the major contribution to the infrared absorption stems from the dark $\Gamma$-hole excitons, where the holes located at the $\Gamma$ point have a large effective mass resulting in a strong excitonic binding energy and consequently a pronounced blue-shift of  $\unit[40]{meV}$ of the 1s-2p transition energy in the incoherent regime. Here, the Coulomb exchange coupling lifts up spin-like states leaving spin-unlike $\Gamma$-K excitons as the lowest states. This has already been confirmed in a recent experiment \cite{Potemski2017a},  where a brightening of these states has been observed in the presence of an in-plane magnetic field. 

In conclusion, we have shown how one can directly map the excitonic landscape in monolayer transition metal dichalcogenides by probing dark and bright exciton populations using infrared spectroscopy. 
We distinguish the optical response shortly after  excitation  stemming from coherent optically excited excitons and the response after  exciton thermalization  reflecting the population of energetically lowest incoherent excitons.  This allows us to follow the exciton dynamics in different regimes and in particular to identify the energetically lowest state. We find that 
MoSe$_2$ is the only direct-gap semiconductor, while MoS$_2$, WS$_2$, and WSe$_2$ exhibit dark excitons that lie below the optically accessible state. Our findings shed light to the fascinating dark excitonic landscape in TMDs and give new insights that will be relevant for the technological exploitation of atomically thin 2D materials.

{\small{
\section{Appendix}
The applied theoretical approach is based on the density matrix formalism \cite{PQE,Kochbuch, malic13} taking into account all relevant many-particle interactions on microscopic footing. 
The considered many-particle Hamilton operator accounts for the Coulomb interaction, carrier-phonon, and carrier-light interaction. All occurring matrix elements are evaluated within an effective Hamiltonian tight-binding approach and explicitly include TMD characteristic symmetries and coupling strengths. The carrier-phonon interaction has been taken into account to calculate the excitonic linewidths as well as the phonon-assisted formation and thermalization of incoherent excitons \cite{Malte2016a,2017_Malte_Arxiv}. The Coulomb interaction is considered within the Keldysh potential taking into account including the substrate-induced screening \cite{Keldysh1979,BerkelbachPRB2013,Berkelbach2015PRB,Bergh2014b,RubioPRB2012}. The carrier-light interaction is accounted for by the light-matter Hamiltonian in the $p\cdot A$ gauge \cite{2008_Steiner_PRB,2008_Steiner_PRL,2010_Steiner_PRL}. It consists of an inter- and an intra-band contribution $H_{\text{c-l}}=\sum_{\lambda} \left(H_{\mathbf{A}\cdot\mathbf{p}}^{\lambda\lambda}+H_{\mathbf{A}\cdot\mathbf{p}}^{\lambda\bar{\lambda}}\right)$. While the first is determined by the expectation value of the momentum operator and has been investigated in previous studies \cite{Bergh2014b,Moody2015,Malte2016a}, we focus here on the intra-band Hamiltonian $H_{\mathbf{A}\cdot\mathbf{p}}^{\lambda\lambda}=\sum_{\lambda}j_{\lambda}(\mathbf k) A(t) \, a_{\lambda\mathbf k}^{\dagger} a_{\lambda\mathbf k}$. Here, $j_{\lambda}(\mathbf k)$ is  the intra-band current projected to the polarization direction $\mathbf e_A$ of the incident light field 
$\mathbf A(t)=A(t)  \,\mathbf e_A$. The current  reads $
 \mathbf j_{\lambda}(\mathbf k)=-\frac{e_0}{\hbar}\frac{\partial \varepsilon_{\mathbf k}^{\lambda}}{\partial \mathbf k}=- \frac{e_0\hbar} {m_{\lambda}} \,\mathbf k
 $ with the effective mass $m_\lambda$ of the considered band $\lambda$ \cite{Kormanyos2014}.

To determine the infrared absorption of TMDs, we calculate the macroscopic intraband current $J(t)$ as defined in Eq. (\ref{eq:MacrointrabandCurrenttime}). It includes the optical response in the coherent regime shortly after the optical excitation as well as in the incoherent regime after the exciton thermalization.
The latter is  determined by the thermalized incoherent exciton density $N_{\mathbf Q}^{n}$ that is approximated by 
 Bose distribution. The coherent response is given by the excitonic microscopic polarization $p^n_{\mathbf Q}(t)$ describing the optical response to a weak probe pulse with the field amplitude $A_0(t)$. It is given by the TMD Bloch equation 
 \begin{align}
\label{eq:dynexc}
 i\hbar\dot{p}_{\mathbf Q}^{n}(t)&=\left(\varepsilon^{n}_{\mathbf Q}-i\gamma_{\mathbf Q}^n\right)\,p_{\mathbf Q}^{n}(t)
 -\sum_{m}j_{nm} A_{0}(t)\, p^{m}_{\mathbf Q}(t)\notag\\&\notag
 -\sum_{\bm q}\Omega^{n}_{\bm q}(t)\bigg(1-2\sum_{lm} \phi_{\mathbf q}^{m}\phi_{\mathbf q}^{l *}\,p^{l}_{\mathbf Q}(t)p^{m*}_{\mathbf Q}(t)\bigg)\\&
 -\Gamma^{n m}_{ij}\,p^{l}_{\mathbf Q}(t)p^{m*}_{\mathbf Q}(t)p_{\mathbf Q}^{n}(t).
 \end{align}
 Here, we have introduced the excitonic interband Rabi frequency
 $\Omega^{n }_{\bm q}(t)= \phi_{\mathbf q}^{n  *}\, \mathbf M^{vc}_{\mathbf q} \cdot  \mathbf A(t)$ with the optical matrix element 
$M^{vc}_{\mathbf q}$ \cite{Bergh2014b}. Furthermore,  it appears the excitonic dispersion $\varepsilon^{n}_{\mathbf Q}=E^{n}+\frac{\hbar^2 Q^2}{2M}$ with the excitonic eigenenergy $E^n$ that is obtained by solving the Wannier equation.
To account for higher-order dephasing of the excitonic polarization, we have added a material and temperature dependent dephasing rate $\gamma^n_{\mathbf Q}$ that has been obtained by calculating exciton-phonon scattering channels in these materials \cite{Malte2016a, Moody2015}. 
 Finally, we  have included the Coulomb-induced non-linear coupling term 
$$
\Gamma^{n m}_{ij} = \sum_{\bm q \bm {q'}}
  \phi_{\mathbf q}^{n *}
\bigg(
2 \phi_{\mathbf q}^{i }
  \phi_{\mathbf q}^{j *}
  \phi_{\mathbf q' }^{m } 
V^{cv \mathbf q\mathbf q'}_{cv \mathbf q'\mathbf q}
- \phi_{\mathbf q' }^{i }
  \phi_{\mathbf q'}^{j *}
  \phi_{\mathbf q }^{m } 
\sum_{\lambda}
V^{\lambda\lambda \mathbf q\mathbf q'}_{\lambda\lambda \mathbf q'\mathbf q}\bigg)
$$
 that gives rise to the excitation-induced renormalization of the Rabi frequency and the electronic band structure.

After solving  the TMD Bloch equation (Eq.(\ref{eq:dynexc})) for the excitonic polarization $p_{\mathbf Q}^{n}(t)$, we have access to the intra-band current from  Eq.(\ref{eq:MacrointrabandCurrenttime}) and can calculate  the optical susceptibility 
  $\chi(\omega)$ determining the optical response to an infrared pulse 
 in the coherent and incoherent regime.

\section{Acknowledgments}
We acknowledge funding from the European Unions Horizon 2020 research and innovation programm 
under grant agreement No 696656 (Graphene Flagship) and from the project 734690 (SONAR). 
Furthermore, the work was supported by the Swedish Research Council (VR) and the Deutsche Forschungsgemeinschaft (DFG) through Research Training Group GK1570 and collaborative research center SFB 1277 (project A05). 
Finally we thank P. Nagler and T. Korn for helpful discussions and experimental support.


\begin{thebibliography}{10}
\expandafter\ifx\csname url\endcsname\relax
  \def\url#1{\texttt{#1}}\fi
\expandafter\ifx\csname urlprefix\endcsname\relax\def\urlprefix{URL }\fi
\providecommand{\bibinfo}[2]{#2}
\providecommand{\eprint}[2][]{\url{#2}}

\bibitem{novoselov05}
\bibinfo{author}{Novoselov, K.~S.} \emph{et~al.}
\newblock \bibinfo{title}{{Two-dimensional gas of massless Dirac fermions in
  graphene}}.
\newblock \emph{\bibinfo{journal}{Nature}} \textbf{\bibinfo{volume}{438}},
  \bibinfo{pages}{197--200} (\bibinfo{year}{2005}).

\bibitem{HeinzPRL2010}
\bibinfo{author}{Mak, K.~F.}, \bibinfo{author}{Lee, C.}, \bibinfo{author}{Hone,
  J.}, \bibinfo{author}{Shan, J.} \& \bibinfo{author}{Heinz, T.~F.}
\newblock \bibinfo{title}{{Atomically Thin MoS$_2$: A New Direct-Gap
  Semiconductor}}.
\newblock \emph{\bibinfo{journal}{Phys. Rev. Lett.}}
  \textbf{\bibinfo{volume}{105}}, \bibinfo{pages}{136805}
  (\bibinfo{year}{2010}).

\bibitem{Splendiani2010}
\bibinfo{author}{Splendiani, A.} \emph{et~al.}
\newblock \bibinfo{title}{{Emerging photoluminescence in monolayer MoS$_2$.}}
\newblock \emph{\bibinfo{journal}{Nano Letters}} \textbf{\bibinfo{volume}{10}},
  \bibinfo{pages}{1271--5} (\bibinfo{year}{2010}).

\bibitem{Chernikov2014}
\bibinfo{author}{Chernikov, A.} \emph{et~al.}
\newblock \bibinfo{title}{{Exciton Binding Energy and Nonhydrogenic Rydberg
  Series in Monolayer WS$_2$}}.
\newblock \emph{\bibinfo{journal}{Physical Review Letters}}
  \textbf{\bibinfo{volume}{113}}, \bibinfo{pages}{76802}
  (\bibinfo{year}{2014}).

\bibitem{Hongkun2017}
\bibinfo{author}{{Zhou You}} \emph{et~al.}
\newblock \bibinfo{title}{{Probing dark excitons in atomically thin
  semiconductors via near-field coupling to surface plasmon polaritons}}.
\newblock \emph{\bibinfo{journal}{Nat Nano}} \textbf{\bibinfo{volume}{advance
  online publication}} (\bibinfo{year}{2017}).

\bibitem{Potemski2017a}
\bibinfo{author}{Molas, M.~R.} \emph{et~al.}
\newblock \bibinfo{title}{Brightening of dark excitons in monolayers of
  semiconducting transition metal dichalcogenides}.
\newblock \emph{\bibinfo{journal}{Scientific Reports}}
  \textbf{\bibinfo{volume}{4}}, \bibinfo{pages}{021003} (\bibinfo{year}{2017}).

\bibitem{2017_Heinz_nnano}
\bibinfo{author}{Zhang, X.-X.} \emph{et~al.}
\newblock \bibinfo{title}{{Magnetic brightening and control of dark excitons in
  monolayer WSe$_2$}}.
\newblock \emph{\bibinfo{journal}{Nat Nano}} \textbf{\bibinfo{volume}{advance
  online publication}} (\bibinfo{year}{2017}).

\bibitem{MacDonald2015PRB}
\bibinfo{author}{Wu, F.}, \bibinfo{author}{Qu, F.} \&
  \bibinfo{author}{MacDonald, A.~H.}
\newblock \bibinfo{title}{{Exciton band structure of monolayer MoS$_{2}$}}.
\newblock \emph{\bibinfo{journal}{Phys. Rev. B}} \textbf{\bibinfo{volume}{91}},
  \bibinfo{pages}{075310} (\bibinfo{year}{2015}).

\bibitem{Berkelbach2015PRB}
\bibinfo{author}{Berkelbach, T.~C.}, \bibinfo{author}{Hybertsen, M.~S.} \&
  \bibinfo{author}{Reichman, D.~R.}
\newblock \bibinfo{title}{Bright and dark singlet excitons via linear and
  two-photon spectroscopy in monolayer transition-metal dichalcogenides}.
\newblock \emph{\bibinfo{journal}{Phys. Rev. B}} \textbf{\bibinfo{volume}{92}},
  \bibinfo{pages}{085413} (\bibinfo{year}{2015}).

\bibitem{Malte2016a}
\bibinfo{author}{Selig, M.} \emph{et~al.}
\newblock \bibinfo{title}{{Excitonic linewidth and coherence lifetime in
  monolayer transition metal dichalcogenides}}.
\newblock \emph{\bibinfo{journal}{Nature Communications}}
  \textbf{\bibinfo{volume}{7}}, \bibinfo{pages}{13279} (\bibinfo{year}{2016}).

\bibitem{Danovich2017a}
\bibinfo{author}{{Danovich Mark}}, \bibinfo{author}{{Z{\'o}lyomi Viktor}} \&
  \bibinfo{author}{{Fal{\rq}ko Vladimir I.}}
\newblock \bibinfo{title}{{Dark trions and biexcitons in WS$_2$ and WSe$_2$
  made bright by e-e scattering}}.
\newblock \emph{\bibinfo{journal}{2D Materials}} \textbf{\bibinfo{volume}{7}},
  \bibinfo{pages}{45998} (\bibinfo{year}{2017}).

\bibitem{2015_Steinleitner_NatMat}
\bibinfo{author}{{Poellmann C.}} \emph{et~al.}
\newblock \bibinfo{title}{{Resonant internal quantum transitions and
  femtosecond radiative decay of excitons in monolayer WSe$_2$}}.
\newblock \emph{\bibinfo{journal}{Nat Mater}} \textbf{\bibinfo{volume}{14}},
  \bibinfo{pages}{889--893} (\bibinfo{year}{2015}).

\bibitem{2015_Heinz_PRL}
\bibinfo{author}{Zhang, X.-X.}, \bibinfo{author}{You, Y.},
  \bibinfo{author}{Zhao, S. Y.~F.} \& \bibinfo{author}{Heinz, T.~F.}
\newblock \bibinfo{title}{{Experimental Evidence for Dark Excitons in Monolayer
  WSe$_{2}$}}.
\newblock \emph{\bibinfo{journal}{Phys. Rev. Lett.}}
  \textbf{\bibinfo{volume}{115}}, \bibinfo{pages}{257403}
  (\bibinfo{year}{2015}).

\bibitem{2017_Feierabend_NatComm}
\bibinfo{author}{Feierabend, M.}, \bibinfo{author}{Bergh{\"a}user, G.},
  \bibinfo{author}{Knorr, A.} \& \bibinfo{author}{Malic, E.}
\newblock \bibinfo{title}{Proposal for dark exciton based chemical sensors}.
\newblock \emph{\bibinfo{journal}{Nature Communications}}
  \bibinfo{pages}{14776}.

\bibitem{2017_Malte_Arxiv}
\bibinfo{author}{Malte, S.} \emph{et~al.}
\newblock \bibinfo{title}{Dark and bright exciton formation, thermalization,
  and photoluminescence in monolayer transition metal dichalcogenides}.
\newblock \emph{\bibinfo{journal}{arXiv:1703.03317}} .

\bibitem{Steinhoff2017a}
\bibinfo{author}{Steinhoff, A.} \emph{et~al.}
\newblock \bibinfo{title}{Excitons versus electron-hole plasma in monolayer
  transition metal dichalcogenide semiconductors}.
\newblock \emph{\bibinfo{journal}{arXiv:1705.05202}} .

\bibitem{Steinhoff2014}
\bibinfo{author}{Steinhoff, A.}, \bibinfo{author}{Rosner, M.},
  \bibinfo{author}{Jahnke, F.}, \bibinfo{author}{Wehling, T.~O.} \&
  \bibinfo{author}{Gies, C.}
\newblock \bibinfo{title}{{Influence of Excited Carriers on the Optical and
  Electronic Properties of MoS$_2$}}.
\newblock \emph{\bibinfo{journal}{Nano Letters}} \textbf{\bibinfo{volume}{14}},
  \bibinfo{pages}{3743--3748} (\bibinfo{year}{2014}).

\bibitem{Darexcitonpaper}
\bibinfo{author}{Malic, E.} \emph{et~al.}
\newblock \bibinfo{title}{in preparation}  (\bibinfo{year}{2017}).

\bibitem{2017_Steinleitner_NanoLetter}
\bibinfo{author}{Steinleitner, P.} \emph{et~al.}
\newblock \bibinfo{title}{{Direct Observation of Ultrafast Exciton Formation in
  a Monolayer of WSe$_2$}}.
\newblock \emph{\bibinfo{journal}{Nano Letters}} \textbf{\bibinfo{volume}{17}},
  \bibinfo{pages}{1455--1460} (\bibinfo{year}{2017}).

\bibitem{2008_Steiner_PRL}
\bibinfo{author}{Lein\ss{}, S.} \emph{et~al.}
\newblock \bibinfo{title}{{Terahertz Coherent Control of Optically Dark
  Paraexcitons in Cu$_{2}$O}}.
\newblock \emph{\bibinfo{journal}{Phys. Rev. Lett.}}
  \textbf{\bibinfo{volume}{101}}, \bibinfo{pages}{246401}
  (\bibinfo{year}{2008}).

\bibitem{2008_Steiner_PRB}
\bibinfo{author}{Steiner, J.~T.}, \bibinfo{author}{Kira, M.} \&
  \bibinfo{author}{Koch, S.~W.}
\newblock \bibinfo{title}{Optical nonlinearities and rabi flopping of an
  exciton population in a semiconductor interacting with strong terahertz
  fields}.
\newblock \emph{\bibinfo{journal}{Phys. Rev. B}} \textbf{\bibinfo{volume}{77}},
  \bibinfo{pages}{165308} (\bibinfo{year}{2008}).

\bibitem{2010_Steiner_PRL}
\bibinfo{author}{Smith, R.~P.} \emph{et~al.}
\newblock \bibinfo{title}{Extraction of many-body configurations from nonlinear
  absorption in semiconductor quantum wells}.
\newblock \emph{\bibinfo{journal}{Phys. Rev. Lett.}}
  \textbf{\bibinfo{volume}{104}}, \bibinfo{pages}{247401}
  (\bibinfo{year}{2010}).

\bibitem{2016_Boland_NNano}
\bibinfo{author}{Joyce, H.~J.}, \bibinfo{author}{Boland, J.~L.},
  \bibinfo{author}{Davies, C.~L.}, \bibinfo{author}{Baig, S.~A.} \&
  \bibinfo{author}{Johnston, M.~B.}
\newblock \bibinfo{title}{{A review of the electrical properties of
  semiconductor nanowires: insights gained from terahertz conductivity
  spectroscopy}}.
\newblock \emph{\bibinfo{journal}{Semiconductor Science and Technology}}
  \textbf{\bibinfo{volume}{31}}, \bibinfo{pages}{103003}
  (\bibinfo{year}{2016}).

\bibitem{2017_Jigang_Nature}
\bibinfo{author}{{Luo Liang}} \emph{et~al.}
\newblock \bibinfo{title}{Ultrafast terahertz snapshots of excitonic rydberg
  states and electronic coherence in an organometal halide perovskite}.
\newblock \emph{\bibinfo{journal}{Nature Communications}}
  \textbf{\bibinfo{volume}{8}}, \bibinfo{pages}{15565} (\bibinfo{year}{2017}).

\bibitem{Kida2005}
\bibinfo{author}{Kida, N.}, \bibinfo{author}{Murakami, H.} \&
  \bibinfo{author}{Tonouchi, M.}
\newblock \emph{\bibinfo{title}{Terahertz Optics in Strongly Correlated
  Electron Systems}}, \bibinfo{pages}{271--330} (\bibinfo{publisher}{Springer
  Berlin Heidelberg}, \bibinfo{address}{Berlin, Heidelberg},
  \bibinfo{year}{2005}).

\bibitem{2007_Masayoshi_NatPhot}
\bibinfo{author}{Tonouchi, M.}
\newblock \bibinfo{title}{Cutting-edge terahertz technology}.
\newblock \emph{\bibinfo{journal}{Nat Photon}} \textbf{\bibinfo{volume}{1}},
  \bibinfo{pages}{97--105} (\bibinfo{year}{2007}).

\bibitem{2016_Hyunyong_NatCom}
\bibinfo{author}{{Cha Soonyoung}} \emph{et~al.}
\newblock \bibinfo{title}{{1s-intraexcitonic dynamics in monolayer MoS$_2$
  probed by ultrafast mid-infrared spectroscopy}} \textbf{\bibinfo{volume}{7}},
  \bibinfo{pages}{10768} (\bibinfo{year}{2016}).

\bibitem{PQE}
\bibinfo{author}{Kira, M.} \& \bibinfo{author}{Koch, S.~W.}
\newblock \emph{\bibinfo{title}{{Many-body correlations and exitonic effects in
  semiconductor spectroscopy}}}, vol.~\bibinfo{volume}{30}
  (\bibinfo{publisher}{IEEE Journal of Quantum Electronics},
  \bibinfo{year}{2006}).

\bibitem{Kochbuch}
\bibinfo{author}{Haug, H.} \& \bibinfo{author}{Koch, S.~W.}
\newblock \emph{\bibinfo{title}{Quantum Theory of the Optical and Electronic
  Properties of Semiconductors}} (\bibinfo{publisher}{World Scientific
  Publishing Co. Pre. Ltd.}, \bibinfo{year}{2004}), \bibinfo{edition}{fifth
  edit} edn.

\bibitem{malic13}
\bibinfo{author}{Malic, E.} \& \bibinfo{author}{Knorr, A.}
\newblock \emph{\bibinfo{title}{Graphene and Carbon Nanotubes: Ultrafast Optics
  and Relaxation Dynamics}} (\bibinfo{publisher}{Wiley-{VCH}},
  \bibinfo{year}{2013}), \bibinfo{edition}{1} edn.

\bibitem{Bergh2014b}
\bibinfo{author}{Bergh{\"{a}}user, G.} \& \bibinfo{author}{Malic, E.}
\newblock \bibinfo{title}{{Analytical approach to excitonic properties of
  MoS$_{2}$}}.
\newblock \emph{\bibinfo{journal}{Phys. Rev. B}} \textbf{\bibinfo{volume}{89}},
  \bibinfo{pages}{125309} (\bibinfo{year}{2014}).

\bibitem{2004_PRL_KochThZGain}
\bibinfo{author}{Kira, M.} \& \bibinfo{author}{Koch, S.~W.}
\newblock \bibinfo{title}{{Exciton-Population Inversion and Terahertz Gain in
  Semiconductors Excited to Resonance}}.
\newblock \emph{\bibinfo{journal}{Phys. Rev. Lett.}}
  \textbf{\bibinfo{volume}{93}}, \bibinfo{pages}{76402} (\bibinfo{year}{2004}).

\bibitem{2016_bergh_2Dmat}
\bibinfo{author}{Bergh\"auser, G.}, \bibinfo{author}{Knorr, A.} \&
  \bibinfo{author}{Malic, E.}
\newblock \bibinfo{title}{Optical fingerprint of dark 2p-states in transition
  metal dichalcogenides}.
\newblock \emph{\bibinfo{journal}{2D Materials}} \textbf{\bibinfo{volume}{4}},
  \bibinfo{pages}{015029} (\bibinfo{year}{2017}).

\bibitem{Kormanyos2014}
\bibinfo{author}{Andor, K.} \emph{et~al.}
\newblock \bibinfo{title}{{k $\cdot$ p theory for two-dimensional transition
  metal dichalcogenide semiconductors}}.
\newblock \emph{\bibinfo{journal}{2D Materials}} \textbf{\bibinfo{volume}{2}},
  \bibinfo{pages}{022001} (\bibinfo{year}{2015}).

\bibitem{Barry_PRL_Louie2015}
\bibinfo{author}{Qiu, D.~Y.}, \bibinfo{author}{Cao, T.} \&
  \bibinfo{author}{Louie, S.~G.}
\newblock \bibinfo{title}{Nonanalyticity, valley quantum phases, and lightlike
  exciton dispersion in monolayer transition metal dichalcogenides: Theory and
  first-principles calculations}.
\newblock \emph{\bibinfo{journal}{Phys. Rev. Lett.}}
  \textbf{\bibinfo{volume}{115}}, \bibinfo{pages}{176801}
  (\bibinfo{year}{2015}).

\bibitem{Echeverry2016PRB}
\bibinfo{author}{Echeverry, J.~P.}, \bibinfo{author}{Urbaszek, B.},
  \bibinfo{author}{Amand, T.}, \bibinfo{author}{Marie, X.} \&
  \bibinfo{author}{Gerber, I.~C.}
\newblock \bibinfo{title}{Splitting between bright and dark excitons in
  transition metal dichalcogenide monolayers}.
\newblock \emph{\bibinfo{journal}{Phys. Rev. B}} \textbf{\bibinfo{volume}{93}},
  \bibinfo{pages}{121107} (\bibinfo{year}{2016}).

\bibitem{2015_Chernikov_PRL}
\bibinfo{author}{Chernikov, A.} \emph{et~al.}
\newblock \bibinfo{title}{{Electrical Tuning of Exciton Binding Energies in
  Monolayer WS$_{2}$}}.
\newblock \emph{\bibinfo{journal}{Phys. Rev. Lett.}}
  \textbf{\bibinfo{volume}{115}}, \bibinfo{pages}{126802}
  (\bibinfo{year}{2015}).

\bibitem{Keldysh1979}
\bibinfo{author}{Keldysh, L.~V.}
\newblock \bibinfo{title}{{Coulomb Interaction in Thin Semiconductor and
  Semimetal Films}}.
\newblock \emph{\bibinfo{journal}{Jetp Letters}} \textbf{\bibinfo{volume}{29}},
  \bibinfo{pages}{658--661} (\bibinfo{year}{1979}).

\bibitem{BerkelbachPRB2013}
\bibinfo{author}{Berkelbach, T.~C.}, \bibinfo{author}{Hybertsen, M.~S.} \&
  \bibinfo{author}{Reichman, D.~R.}
\newblock \bibinfo{title}{Theory of neutral and charged excitons in monolayer
  transition metal dichalcogenides}.
\newblock \emph{\bibinfo{journal}{Phys. Rev. B}} \textbf{\bibinfo{volume}{88}},
  \bibinfo{pages}{045318} (\bibinfo{year}{2013}).

\bibitem{RubioPRB2012}
\bibinfo{author}{Cudazzo, P.}, \bibinfo{author}{Gatti, M.} \&
  \bibinfo{author}{Rubio, A.}
\newblock \bibinfo{title}{Plasmon dispersion in layered transition-metal
  dichalcogenides}.
\newblock \emph{\bibinfo{journal}{Phys. Rev. B}} \textbf{\bibinfo{volume}{86}},
  \bibinfo{pages}{075121} (\bibinfo{year}{2012}).

\bibitem{Moody2015}
\bibinfo{author}{Moody, G.} \emph{et~al.}
\newblock \bibinfo{title}{{Intrinsic Exciton Linewidth in Monolayer Transition
  Metal Dichalcogenides}}.
\newblock \emph{\bibinfo{journal}{Nature Commun.}}
  \textbf{\bibinfo{volume}{6}}, \bibinfo{pages}{8315} (\bibinfo{year}{2015}).

\end{thebibliography}

\end{document}